\documentclass[journal]{IEEEtran}
\usepackage[T1]{fontenc}
\usepackage{amsmath}
\usepackage{amssymb}
\usepackage{graphicx}
\usepackage{cite}
\usepackage{booktabs}
\usepackage{url}
\usepackage{microtype}
\usepackage{array}
\usepackage{multirow}
\usepackage[hidelinks]{hyperref}

\begin{document}

\title{Multi-Domain Physics-Based MDO of Multirotor UAVs: \\
A Deterministic Framework for Discrete COTS Sizing}

\author{
  \textbf{Akshay Gupta Burela}\\
  {\small Department of Electronics and Communication Engineering,\\
  International Institute of Information Technology, Hyderabad, India\\
  Email: akshay.burela@research.iiit.ac.in}\\[1.5ex]
  \textbf{P. B. Sujit}\\
  {\small Department of Electrical Engineering and Computer Science,\\
  Indian Institute of Science Education and Research, Bhopal, India\\
  Email: sujit@iiserb.ac.in}
}

\markboth{\MakeLowercase{ar}X\MakeLowercase{iv} P\MakeLowercase{reprint 2026 --- }A\MakeLowercase{ero}E\MakeLowercase{val}}%
{A\MakeLowercase{ero}E\MakeLowercase{val}: P\MakeLowercase{hysics-}B\MakeLowercase{ased }COTS M\MakeLowercase{ultirotor }S\MakeLowercase{izing}}

\maketitle

\begin{abstract}
Multirotor Unmanned Aerial Vehicle (UAV) design is governed by a
tightly coupled system of non-linear equations spanning structural
mechanics, electrochemistry, aerodynamics, and kinematics.  Solved
sequentially, a miscalibrated sub-model coefficient triggers a
mass-compounding cascade failure---the \emph{Mass Snowball} effect.
This paper presents \textbf{AeroEval}, a physics-based,
Multidisciplinary Design Optimization (MDO) engine that simultaneously
resolves all subsystem couplings and maps continuous sizing optima to
physically purchasable, commercial off-the-shelf (COTS) components. 


The MDO engine is validated under a strict calibrate/test protocol that
eliminates circularity: three structural/packaging coefficients are
fitted on a held-out cohort of 20 do-it-yourself (DIY) and legacy
platforms, then frozen and evaluated \emph{blind} on 19 modern
commercial drone platforms spanning 377\,g to 76\,kg .  On the commercial test cohort the engine predicts Maximum
Takeoff Weight (MTOW) within \textbf{7.2\% Mean Absolute Percentage
Error (MAPE)}, with a mean complexity factor
$k \approx 1.05 \pm 0.08$ and $\text{RMSE}_{\text{MTOW}} = 1.59$\,kg;
the DIY calibration cohort, characterized by high build variability,
yields 26.1\% MAPE.  Battery mass is predicted within 7.9\% on
single-pack platforms, with a disclosed systematic underprediction on
redundant multi-battery enterprise platforms.  A 14-parameter
sensitivity suite of $>$300 simulations quantifies partial derivatives
of takeoff mass and flight time; forward velocity diverges beyond
25.5\,m/s due to cubic parasite power growth.  The path-dependent
mass-shedding model for agricultural and delivery roles reduces
structural frame mass by up to 40.6\% and energy capacity by 33.7\%
relative to static baselines.  Typical solver convergence requires
25--50 iterations, completing in under 50\,ms on a standard desktop
CPU.

\end{abstract}


\section{Introduction}
\label{sec:intro}
\IEEEPARstart{M}{ultirotor} unmanned aerial vehicles constructed from
commercial off-the-shelf (COTS) components are deployed across
cinematography, precision agriculture, last-mile delivery, bridge
inspection, and emergency response.  Despite their ubiquity, the
design of a multirotor for a specified payload and mission remains a
computationally challenging inverse problem.

Structural frame mass, battery pack capacity, motor torque constants,
and rotor aerodynamic performance are mutually dependent: a heavier
structure demands a larger battery; a larger battery demands
higher-torque motors; larger motors require larger propellers; larger
propellers impose greater structural loads.  Sequential solvers that
handle these sub-problems in isolation diverge to infinite takeoff
weight when any individual sub-model is slightly miscalibrated---the
Mass Snowball effect.

\subsection{Limitations of Existing Tools}
eCalc~\cite{ecalc} provides a browser-based static estimator that
ignores two-way structural--aerodynamic coupling and performs no
transient thermal simulation.  OpenVSP~\cite{openvsp} targets
fixed-wing geometry; its rotor modules lack electrochemical discharge
modeling.  SUAVE~\cite{suave} solves the multidisciplinary problem but
is oriented toward large Vertical Takeoff and Landing (VTOL) platforms
and requires calibrated component databases unavailable for hobbyist
COTS parts.  QGroundControl~\cite{qgc} manages flight operations but
performs no sizing.  None of these tools simultaneously models
transient battery thermodynamics, structural resonance verification,
and COTS hardware database search under a formal convergence guarantee.

\subsection{Contributions}
AeroEval addresses these gaps with the following contributions:
\begin{itemize}
  \item \textbf{Convergent MDO solver} --- A Banach fixed-point
        mass-iteration loop with formal convergence guarantees and
        dynamic under-relaxation, valid for any design where the
        mass-growth Lipschitz constant $L < 1$.
  \item \textbf{Scale-invariant electrochemical--thermal battery model}
        --- A thermal model with capacity-proportional convective
        cooling ($hA_c$) spanning a $130\times$ dissipation range,
        validated across vehicle masses from 377\,g to 76\,kg.  It
        features twelve battery chemistry profiles using a
        Tremblay--Dessaint Open-Circuit Voltage (OCV) formulation with
        C-rate scaling and cycle-aging limits.
  \item \textbf{Path-dependent mass-shedding} --- Time-step integrator
        for agricultural continuous spray and delivery discrete drop
        roles, reducing sizing conservatism by up to 33.7\%.
  \item \textbf{Non-circular empirical validation} --- A calibrate/test
        protocol: three structural/packaging coefficients are fitted on
        a held-out 20-platform DIY cohort, frozen, then evaluated blind
        on 19 commercial platforms, with quantified $k$-factors and
        scope boundaries.
  \item \textbf{Dual-mode accessible interface} --- A requirements-first
        Mission Workspace layer for non-expert practitioners and a full
        57-parameter Physics Workspace for engineers, making
        physics-rigorous UAV sizing available across the entire
        practitioner spectrum without simplifying the underlying physics.
  \item \textbf{Declarative hardware lock and override system} --- Any
        combination of battery, motor, and propeller hardware can be
        locked as a hard constraint; the solver re-sizes all remaining
        subsystems around the constraint, functioning as a
        constraint-satisfaction advisor for operators already owning
        specific components.
  \item \textbf{Runtime-editable COTS database} --- Three JSON libraries
        (motors/propellers, battery chemistries, ESCs) loaded at runtime
        without recompilation, enabling immediate integration of new
        market components and custom proprietary hardware.
  \item \textbf{Dual-run comparative analysis} --- Each solver
        invocation executes two independent pipeline runs: a
        null-role \emph{baseline} (dead-weight hover math, thrust-to-weight
        (T/W) ratio $\approx 2.78$) and a \emph{custom} run with the full
        active role applied, isolating pure-aerodynamic MTOW from
        role-induced overhead on every invocation.
  \item \textbf{Validated numerical robustness} --- A 102-test
        verification suite (28 physics/monotonicity checks, 74
        full-schema coverage tests) operates across all 12 chemistry
        profiles, 6 role types, 4 airframe classes, and all lock and
        override combinations; no unexpected failures under the
        calibrated model were observed across $>$500 engine invocations.
\end{itemize}

\section{System Architecture and Design Philosophy}
\label{sec:arch}

\subsection{Design Philosophy}
AeroEval applies a \emph{conservative lower-bound} MDO philosophy.
The engine sizes around standard Glauert momentum theory, explicit
aerospace safety factors, and COTS parts.  A positive safety margin is
targeted: if the engine accepts a design, it is intended to fly in
practice.

The engine operates on a \emph{COTS-first} principle.  Continuous
optimization routinely yields theoretical optima (e.g., a
14.37-inch propeller at $KV = 341.25$\,rpm/V) that cannot be
purchased.  AeroEval uses the continuous optimum as coordinates in
the COTS database search space, ensuring every output is immediately
buildable.

\subsection{Five-Module Sequential Pipeline}
\label{ssec:pipeline}
The solver executes five modules in a tightly coupled loop:
\begin{enumerate}
  \item \textbf{Feasibility Gate} --- Low-cost geometric and physical
        checks rejecting impossible configurations.
  \item \textbf{Mission Profiler / Kinematics} --- ISA atmospheric
        model, three-phase mission profiling, aerodynamic drag,
        pitch kinematics, and auto-velocity sweep.
  \item \textbf{Mass Iteration Solver} --- Banach fixed-point loop
        coupling battery sizing, structural arm sizing, motor
        selection, and frame mass.
  \item \textbf{Thermodynamic Discharge Simulator} --- Time-step
        integration of the Tremblay--Dessaint OCV model with dynamic
        internal resistance, battery cell temperature, and motor
        winding heat.
  \item \textbf{Discrete COTS Optimizer} --- Exhaustive guided grid
        search over the propeller-motor-battery COTS database.
\end{enumerate}

\subsection{Software Architecture and Implementation}
\label{ssec:impl}
AeroEval is available as an open-source repository at
\url{https://github.com/AKSHAY-RSOL/AeroEval}.  The directory
structure is summarized in Table~\ref{tab:repodir}.  The solver is
implemented as a self-contained, cross-platform C++17 executable with
no runtime dependencies beyond the C++ standard library; the full
solver---spanning all physics, discrete database search, and output
formatting---is compiled into a single binary.

\begin{table}[ht]
\caption{AeroEval Repository Directory Structure.}
\label{tab:repodir}
\centering
\scriptsize
\begin{tabular}{lp{6.0cm}}
\toprule
Directory/File Path & Primary Purpose and Key Contents \\
\midrule
\texttt{CMakeLists.txt} & Build configuration for compiling the C++ executable. \\
\texttt{README.md} & Compilation instructions, C++ engine commands, and Next.js setup guide. \\
\texttt{data/} & Electrochemical chemistries, ESC database, COTS propeller/motor specifications, and mission example JSONs. \\
\texttt{include/} & C++ header files organized into functional modules (Feasibility, Kinematics, Physics, Optimizer, Validation, Roles, Utils). \\
\texttt{src/} & C++ source files matching the module headers, and the main process coordinator. \\
\texttt{tests/} & Python validation suite and automated parameter sweeps. \\
\texttt{web\_ui/} & Next.js frontend (dashboard page, server actions, styling, configuration). \\
\bottomrule
\end{tabular}
\end{table}

\subsubsection{C++ Module Structure}
The source tree comprises eight purpose-specific modules:
\texttt{main.cpp} (entry point, JSON parsing, dual-run orchestration);
\texttt{MassIteration.cpp} (fixed-point loop, three-phase profiling,
ISA model, altitude-corrected Banach seed);
\texttt{Thermodynamics.cpp} (Tremblay--Dessaint OCV, transient thermal
simulation, voltage-collapse detection, power-overflow guard,
cycle-aging);
\texttt{Structures.cpp} (Euler--Bernoulli arm sizing, cantilever
resonance verification, CG-offset analysis, minimum wall-thickness
floor);
\texttt{DiscreteMINLP.cpp} (guided exhaustive grid search with
multi-objective $J$-score ranking);
\texttt{TaxonomyRoles.hpp} (mission role taxonomy with per-role thrust
margins, load factors, drag penalties, auxiliary power schedules);
\texttt{Aerodynamics.cpp} (Reynolds-scaled Figure of Merit, Glauert
Newton--Raphson induced-velocity solver, rotor-arm interference); and
\texttt{FeasibilityGate.cpp} (pre-solve geometric and physical
constraint validation).

\subsubsection{Structured JSON Input/Output Contract}
All solver parameters and results pass through a strict JSON file
contract, decoupling the physics engine from any caller layer.  The
input schema defines 57 named fields spanning physical bounds, mission
phase parameters, battery chemistry selection, hardware lock
declarations, and physics-constant override flags.  All fields carry
documented defaults; a caller specifying only \texttt{payload\_kg},
\texttt{max\_diameter\_m}, and a mission duration receives a fully
converged result.  The output schema returns MTOW constituent
breakdown, selected COTS components, per-phase energy budgets, thermal
diagnostics, convergence iteration count, and explicit
constraint-violation reason strings.

\subsubsection{Dual-Run Comparative Analysis}
\label{ssec:dualrun}
Each solver invocation executes two independent pipeline runs over the
same input.  The \textbf{baseline run} sets the active role pointer to
\texttt{null} and applies pure dead-weight hover math with no
role-specific drag, auxiliary power, or thrust margin; the baseline
hover throttle target defaults to 60\%, yielding
$\text{T/W} = 1/0.60^2 \approx 2.78$ universally, and represents the
minimum physically possible MTOW for the given aerodynamic
configuration.  The \textbf{custom run} passes the full active role
object through the pipeline, applying role-specific thrust margins,
dynamic load factors, drag multipliers, auxiliary power draws, and
mass-shedding integrators; the T/W ratio reflects the role's actual
hover throttle target, $\text{T/W} = 1/\theta_{\text{hvr}}^2$.  The
difference between custom and baseline MTOW directly quantifies the
overhead imposed by the selected role.

\subsubsection{Hardware Lock and Override System}
A key software contribution is the declarative hardware lock and
override system.  The engine accepts \textbf{locked} declarations for
any combination of: battery capacity and cell count, system bus
voltage (series cells), propeller diameter and pitch, and motor KV
rating.  When any lock is active, the corresponding variable is treated
as a non-iterable constant, and the solver re-sizes all remaining
degrees of freedom around the constraint.  This transforms the engine
from a pure synthesis tool into a constraint-satisfaction advisor---a
workflow absent from all current alternatives reviewed in
Section~\ref{sec:intro}.  Beyond hardware locks, 19 physics-constant
overrides are exposed for expert users possessing empirical test data,
including Figure of Merit (installed/isolated mode), propulsive
efficiency, horizontal and vertical drag coefficients, frontal-area
scaling exponent, aerodynamic body class ($C_d A = 0.004$--$0.05$),
thrust coefficient $C_t$, propeller material class, CFRP tube geometry
constant, wall-thickness ratio, body mass multiplier, and arm
configuration.

\subsubsection{Runtime-Editable COTS Database}
The hardware database is structured as three independently editable
JSON libraries loaded at process startup: \texttt{cots\_database.json}
(motor KV, mass, maximum current; propeller diameter/pitch
combinations), \texttt{chemistry\_profiles.json} (Tremblay--Dessaint
parameters and specific energy per chemistry), and an ESC voltage/current
ratings table.  Database entries are not compiled into the binary,
permitting new components---including proprietary OEM parts---to be
added without recompilation.

\subsection{Banach Fixed-Point Convergence}
\label{ssec:banach}
The mass-compounding loop is formalized as a contractive mapping
$f: X \to X$ on the complete metric space $(X, d)$,
$X \subset \mathbb{R}^+$, $d(x_1,x_2) = |x_1 - x_2|$.  The implicit
mass-balance equation is:
\begin{equation}
M = f(M) = M_{\text{payload}} + M_{\text{bat}}(M) + M_{\text{frame}}(M) + M_{\text{prop/mot}}(M).
\label{eq:implicit}
\end{equation}
Convergence is guaranteed when the mass-growth Lipschitz constant
satisfies:
\begin{equation}
L \approx \frac{\partial f(M)}{\partial M}
= \frac{\partial M_{\text{bat}}}{\partial M}
+ \frac{\partial M_{\text{frame}}}{\partial M}
+ \frac{\partial M_{\text{prop/mot}}}{\partial M} < 1.
\label{eq:lipschitz}
\end{equation}
If $L \geq 1$, every additional kilogram of takeoff mass demands more
than one kilogram of subsystem mass---physical divergence.

The fixed-point formulation offers a deterministic alternative to MDO
approaches that solve the mass loop via heuristic optimization.
Delbecq et~al.~\cite{delbecq2020} formulated an identical mass-coupling
equation but solved it using a Normalized Variable Hybrid MDO
formulation with differential evolution, requiring substantially more
function evaluations than a direct fixed-point iteration.  Unlike MDO
formulations that guarantee neither global optimality nor
convergence~\cite{hosseini2020}, the Banach approach provides a formal
convergence certificate: if $L < 1$, convergence is guaranteed
regardless of initial seed, and the error bound decreases geometrically
as $|M^{(k)} - M^*| \leq L^k |M^{(0)} - M^*|$~\cite{kreyszig1978},
where $M^{(k)}$ is the estimated takeoff mass at iteration $k$,
$M^{(0)}$ the initial seed, $M^*$ the unique converged fixed point, and
$L$ the mass-growth Lipschitz constant.

Numerical step-discontinuities (integer cell-count rounding, discrete
database lookups) cause limit-cycle oscillations even when $L < 1$.
Under-relaxation is applied to the battery capacity estimate as a
numerical damping mechanism:
\begin{equation}
C^{(k+1)} = \alpha\,C^{(k)}_{\text{target}} + (1-\alpha)\,C^{(k)},
\quad \alpha = 0.5,
\label{eq:relaxation}
\end{equation}
where $C^{(k)}_{\text{target}}$ is the thermodynamically required
capacity at iteration $k$ scaled by a 20\% safety margin.  A ratchet
floor $C^{(k+1)} \geq C_{\text{max-failed}} \times 1.05$ prevents
regression to previously infeasible capacity states.  Convergence is
declared when $|M^{(k+1)} - M^{(k)}| < 0.001$\,kg.  If $M > 150$\,kg, a
runaway-divergence flag is set and the solver halts.  When the design
cannot sustain the requested flight duration, an automatic goal
relaxation scales down mission time,
$t_{\text{active}} = \gamma\,t_{\text{requested}}$, with the solver
aborting if $\gamma < 0.05$.  Typical convergence requires 25--50
iterations, completing in $<50$\,ms.

\subsubsection{Altitude-Corrected Banach Seed}
\label{ssec:altcorrect}
A known pathology of fixed-point mass solvers is convergence to
\emph{local attractors} when the initial seed is far from the true
solution.  At high altitude, lower air density increases induced hover
power as $P \propto \rho^{-1/2}$, requiring a heavier battery.  AeroEval
scales the initial mass seed:
\begin{equation}
M^{(0)} = \frac{M_{\text{payload}}}{0.60}
           \cdot \sqrt{\frac{\rho_{\text{SL}}}{\rho(h)}},
\quad \rho_{\text{SL}} = 1.225\,\text{kg/m}^3,
\label{eq:altcorrect}
\end{equation}
ensuring the solver starts from an altitude-appropriate estimate.  The
correction has no effect at sea level and is applied only when the
scale factor exceeds 1.0.  Empirical validation confirms that this
correction substantially improves MTOW monotonicity from sea level to
5000\,m relative to an uncorrected seed, though it does not eliminate
all local non-monotonicity (Section~\ref{sec:limits}).

\subsection{Mission Role Taxonomy}
\label{ssec:roles}
Six mission role profiles parametrize thrust margins, dynamic load
factors, drag penalties, and auxiliary power draws
(Table~\ref{tab:roles}).  The thrust margin $\theta_{\text{margin}}$
sets the hover throttle target
$\theta_{\text{hvr}} = 1 - \theta_{\text{margin}}$, which determines
$\text{T/W} = 1/\theta_{\text{hvr}}^2$.

\begin{table}[ht]
\caption{Mission Role Taxonomy Key Parameters (from \texttt{TaxonomyRoles.hpp}).}
\label{tab:roles}
\centering
\scriptsize
\setlength{\tabcolsep}{2.5pt}
\begin{tabular}{lcccp{3.0cm}}
\toprule
Role & \shortstack{Thrust\\Margin} & \shortstack{Load\\Factor} & \shortstack{Aux. Power\\(W)} & Special \\
\midrule
Imaging       & 45\%  & 1.5\,G  & 15--30 & gimbal drag $0.02\,\text{m}^2$ \\
Delivery      & 40\%  & 1.5\,G  & 5--10  & discrete drop at $t_{\text{drop}}$ \\
Agriculture   & 40\%  & 1.5\,G  & 40--200 & continuous spray; downwash cap \\
Racing        & \textbf{80\%} & \textbf{10.0\,G} & 0--5 & continuous $\text{T/W} \approx 25$ target \\
Inspection    & \textbf{50\%} & 1.5\,G & 20--50 & wall-turbulence gust margin \\
Mapping       & 40\%  & 1.5\,G  & 50--80 & $1.5\times$ LiDAR drag multiplier \\
\bottomrule
\end{tabular}
\end{table}

The racing role's 80\% thrust margin ($\theta_{\text{hvr}} = 0.20$)
yields a continuous design target $\text{T/W} = 25.0$, consistent with
First-Person View (FPV) freestyle quads operating at 20\% hover
throttle.  In discrete COTS matching the selected hardware provides a
validated maximum static $\text{T/W} \approx 9.8$--$13.8$ depending on
motor availability under G-load constraints.  Any margin can be
overridden per-call via \texttt{thrust\_margin\_override}.

\section{User Interface and Accessibility Design}
\label{sec:ui}
A primary objective of the AeroEval software system is to make
physics-rigorous UAV sizing accessible across a broad expertise
spectrum---from students and hobbyists to professional aerospace
engineers.  Existing tools force a binary choice: simplified consumer
tools hide physics behind opaque lookup tables, while research-grade
frameworks demand graduate-level expertise to configure.  AeroEval
exposes the same converged physics result through two qualitatively
different interaction modes without duplicating or compromising the
underlying solver.

\subsection{Mission Workspace: Requirements-First Design}
\label{ssec:kmode}
The Mission Workspace abstracts the 57-parameter input schema behind a
three-dimensional requirements specification: (1)~\textbf{payload mass}
via annotated slider or preset buttons (``DSLR Camera 1.5\,kg'',
``Survey LiDAR 2.0\,kg'', ``10\,L Spray Tank''); (2)~a
\textbf{transport constraint} footprint category (backpack, car trunk,
industrial) that internally sets the wheelbase ceiling $W_{\max}$ and
triggers coaxial layout logic; and (3)~a \textbf{mission role and
optimization priority} selected from six role cards and four priority
axes (maximum endurance, range, payload, or minimum system mass).  From
these three signals, an intent-to-parameter mapping layer sets
chemistry heuristics, thrust margin, airframe class, and auxiliary
power before calling the solver.  A \emph{live design rationale panel}
renders plain-language explanations of each solver decision (e.g.,
``Sized as an Octocopter in Coaxial X8 layout\ldots''), making the
engine's reasoning legible to users who cannot interpret the underlying
equations.

\subsection{Physics Workspace: Full Parameter Control}
\label{ssec:htw}
Engineers may switch to the Physics Workspace, which renders all 57
input parameters across seven logical panels: Physical Bounds; Mission
Profile; Hardware Constraints and Locks (twelve chemistry profiles and
four declarative lock channels); Taxonomy and Role-Specific Math
Overrides; Advanced Scientific Overrides (all 19 physics constants with
academic default annotations); Marketing Spec Simulation (an optional
mode reproducing manufacturer best-case endurance figures); and a
Database Customizer (a three-tab runtime editor for the motor/propeller
library, chemistry table, and ESC catalog).  Every constant that
affects the result is visible, annotated, and overridable---no ``magic
numbers'' are hardcoded invisibly.

\subsection{Shareable Build State and Results Visualization}
Any complete configuration---all 57 parameter values, lock states,
chemistry selections, override flags, and role settings---can be
serialized to a compact base64-encoded string appended to the
application URL, supporting reproducible sharing between design teams,
instructor-to-student distribution, and client-to-manufacturer
handoff.  The results layer presents four interactive components: an
annotated \textbf{mass budget breakdown}; a Canvas-rendered \textbf{COTS
motor crosshair plot} positioning every database motor relative to the
theoretical optimum $(KV_{\text{ideal}}, M_{\text{ideal}})$ with
per-motor $J$-scores; a color-coded SVG \textbf{thermal state
schematic} (green $<40^\circ$C, amber $40$--$80^\circ$C, red
$>80^\circ$C); and a \textbf{parametric endurance explorer} plotting
flight time against a battery capacity scale factor with the
mass-growth penalty inflection marked.

\section{Aerodynamic and Kinematic Modeling}
\label{sec:aero}

\subsection{Feasibility Gate}
\subsubsection{Propeller Overlap Constraint}
The center-to-center distance between adjacent motor hubs follows from
the law of cosines,
$d_{\text{adj}} = 2R_{\text{arm}}\sin(\pi/N_r) = W\sin(\pi/N_r)$,
with $R_{\text{arm}} = W/2$.  Including a tip-clearance safety factor
$\varepsilon = 0.05$:
\begin{equation}
D_p \leq (1-\varepsilon)\,W\sin\!\left(\frac{\pi}{N_r}\right).
\label{eq:overlap}
\end{equation}
The geometric maxima (no clearance) are Quad $D_p \leq 0.707\,W$, Hex
$\leq 0.500\,W$, Octo $\leq 0.383\,W$; with $\varepsilon = 0.05$ these
scale to $0.672\,W$, $0.475\,W$, and $0.364\,W$.  For coaxial layouts
($N_{\text{arms}} = N_r/2$), $N_r$ is replaced by $N_{\text{arms}}$.

\subsubsection{Disk Loading and Downwash}
The hover induced velocity per rotor is Froude's actuator disk
$v_i = \sqrt{T/(2\rho A_s)}$, $A_s = \pi D_p^2/4$.  The total fleet
downwash accounting for all $N_r$ rotors is:
\begin{equation}
v_{\text{dw}} = 2v_i = 2\sqrt{\frac{M_{\text{to}}\,g}{2\rho N_r A_s}}.
\label{eq:downwash}
\end{equation}
For agricultural spraying, $v_{\text{dw}} \leq 8.0$\,m/s prevents crop
damage, vortex-ring-state recirculation, and spray drift.

\subsection{ISA Atmospheric Model}
\label{ssec:isa}
Tropospheric layer ($h \leq 11{,}000$\,m):
\begin{equation}
\rho(h) = \frac{P(h)}{R\,T_{\text{ambient}}},\quad
P(h) = P_0\!\left(1 - \frac{L_r h}{T_0}\right)^{g/(R L_r)},
\label{eq:trop}
\end{equation}
with $P_0 = 101{,}325$\,Pa, $T_0 = 288.15$\,K,
$g = 9.80665\,\text{m/s}^2$, $R = 287.058\,\text{J/(kg K)}$,
$L_r = 0.0065$\,K/m, exponent $\approx 5.256$~\cite{isa1993}.  In the
stratosphere ($11{,}000 \leq h \leq 20{,}000$\,m), temperature is
isothermal at $T_{11} = 216.65$\,K and
$P(h) = P_{11}\exp[-g(h-11{,}000)/(R\,T_{11})]$.

\subsection{Three-Phase Mission Profile and Power Budget}
\label{ssec:threephase}
The mission comprises hover, climb, and cruise phases.  Thrust per
rotor is $T_{\text{hover}} = M_{\text{to}}\,g/N_r$ in hover and
$T_{\text{climb}} = (M_{\text{to}}\,g + F_{d,\text{vert}})/N_r$ in
climb.  Total electrical power in phase $j$ decomposes as:
\begin{equation}
P_{e,j} = \frac{P_{\text{mech},j}}{\eta_{\text{motor}}}
         + \tfrac{1}{2}f_{\text{pwm}}C_{\text{oss}}V_{\text{oc}}^2
         + P_{\text{aux}},
\label{eq:powerbudget}
\end{equation}
with $f_{\text{pwm}} = 24{,}000$\,Hz the PWM switching frequency and
$C_{\text{oss}}$ the MOSFET output capacitance.  Total mission energy
is $E_{\text{req}} = \sum_j P_{e,j}\,t_j/3600$~[Wh], and battery mass
follows as:
\begin{equation}
M_{\text{battery}} = \frac{E_{\text{req}}}{e_{\text{spec}}}\,\eta_{\text{pkg}},
\label{eq:batmass}
\end{equation}
where $e_{\text{spec}}$ is the chemistry specific energy (Wh/kg,
Table~\ref{tab:chem}) and $\eta_{\text{pkg}}$ is a chemistry- and
class-dependent packaging multiplier (Section~\ref{ssec:pkg}).

\subsection{Aerodynamic Drag and Tilted Kinematics}
Parasite drag at cruise $V$ is
$F_d = \tfrac12\rho V^2 C_d S_{\text{ref}}$, with
$S_{\text{ref}} = f_{\text{scale}}M_{\text{to}}^{2/3} + \Delta A_{\text{drag}}$
and $f_{\text{scale}} \in \{0.004, 0.012, 0.05\}$ for compact, bulky,
and research-exposed airframes.  Mapping-role LiDAR payloads apply a
$1.5\times$ drag-area multiplier in forward-flight phases only.
Forward flight requires body pitch
$\theta = \arctan(F_d/(M_{\text{to}}\,g))$ and thrust
$T_{\text{fwd}} = \sqrt{(M_{\text{to}}\,g)^2 + F_d^2}$.

\subsection{Auto-Velocity Optimization}
\label{ssec:autov}
When enabled, the solver sweeps $V \in [1.0, 30.0]$\,m/s in 0.5\,m/s
increments to find $V_{\text{md}} = \arg\min_V P_{\text{fwd}}(V)$.
Sensitivity analysis shows $P_{\text{fwd}}$ diverges for
$V > 25.5$\,m/s on standard Class~1 configurations.

\subsection{Rotor Power Modeling}
Profile power per rotor~\cite{bohorquez2009} is
$P_o = \tfrac18\sigma C_{d,0}\rho A_s V_{\text{tip}}^3$, with
$\sigma = 0.28/\pi$ and $V_{\text{tip}} = \omega R$, and the
Reynolds-dependent drag coefficient:
\begin{equation}
C_{d,0} = \begin{cases}
0.012(10^5/Re_{\text{tip}})^{0.5} & Re_{\text{tip}} < 10^5,\\[3pt]
0.012(10^5/Re_{\text{tip}})^{0.2} & Re_{\text{tip}} \geq 10^5.
\end{cases}
\label{eq:cdre}
\end{equation}
Shaft power is $P_{\text{shaft}} = P_i/\text{FoM}$ with induced power
$P_i = T_{\text{motor}}v_i$ (hover) or $T_{\text{motor}}(V_c+v_i)$
(climb).  In forward flight, Glauert's implicit relation
$v_i = T_{\text{motor}}/(2\rho A_s\sqrt{V_\infty^2 + v_i^2})$ is solved
by Newton--Raphson.  The installed Figure of Merit is reduced by a
class-specific penalty $\kappa$ relative to the isolated lab value,
$\text{FoM}_{\text{installed}} = \text{FoM}_{\text{isolated}}/\kappa$,
with $\kappa \in \{1.20_{\text{consumer}}, 1.25_{\text{enterprise}},
1.30_{\text{MicroAIO}}, 1.46_{\text{industrial}}\}$, and further scaled
by the Bohorquez--Pines Reynolds fit
$\text{FoM}(Re) = \text{FoM}_{\text{max}}(1 - e^{-Re/Re_{\text{ref}}})$,
$Re_{\text{ref}} = 20{,}000$, clamped to a minimum of 0.38.  The
rotor-arm interference correction~\cite{shukla2018} is
$k_{\text{int}} = 1.0 + c\,w_a/D_p$ (with $c = 0.6$ offset, $1.5$
under-rotor, $0$ folding), capped at 1.15.

\section{Structural Sizing and Cantilever Resonance}
\label{sec:struct}

\subsection{Euler--Bernoulli Arm Sizing}
\label{ssec:struct-sizing}
Arms are hollow circular CFRP cantilever tubes~\cite{timoshenko1970} of
length $L_{\text{arm}} = W/2$.  The root bending moment is
$M_b = F_{\text{arm}}L_{\text{arm}}$ with
$F_{\text{arm}} = T_{\text{motor}}\cdot\text{load\_factor}$.  Under a
payload CG offset $x_{\text{cg}}$,
$T_{\text{motor}} = (M_{\text{to}}g/N_r)(1 + x_{\text{cg}}/d_y)$ with
$d_y = L_{\text{arm}}\cos(\pi/N_r)$; if $x_{\text{cg}} \geq d_y$ the
design is rejected.  Imposing $\sigma \leq \sigma_{\text{all}}$ and
solving for the outer diameter:
\begin{equation}
d_{\text{out}} = \sqrt[3]{\frac{32\,M_b}{\pi(1-\alpha_w^4)\sigma_{\text{all}}}},
\quad \alpha_w = 0.90.
\label{eq:dout}
\end{equation}
The fatigue-and-FoS-adjusted allowable bending stress is derived
transparently from the CFRP ultimate strength rather than used raw:
\begin{equation}
\sigma_{\text{all}} = \frac{\sigma_{\text{ult}}\,f_{\text{fat}}}{\text{FoS}}
= \frac{720\times 0.40}{1.5} = 192\,\text{MPa},
\label{eq:sigallow}
\end{equation}
with $\sigma_{\text{ult}} = 720$\,MPa, fatigue knockdown
$f_{\text{fat}} = 0.40$, and $\text{FoS} = 1.5$.  The total carbon-fiber
arm mass from the sized tube dimensions is
$M_{\text{arms}} = N_r\rho_{\text{CF}}\tfrac\pi4(d_{\text{out}}^2 -
d_{\text{in}}^2)L_{\text{arm}}$, with
$\rho_{\text{CF}} = 1600\,\text{kg/m}^3$.  A size-dependent minimum
structural mass fraction
$f_{\text{struct,min}} = \max(0.06,\min(0.20, 0.05 + 0.05\,W))$ is
imposed, and the effective arm mass is
$M_{\text{arms,eff}} = \max(M_{\text{arms}},
M_{\text{to}}f_{\text{struct,min}}/C_{\text{body}})$ with
$C_{\text{body}} = 1.5$.  The central body plate, landing gear, and
mounting mass are then added through a class-specific structural
overhead multiplier $\gamma$ (0.15 ConsumerFolding, 0.50
EnterpriseRugged, 1.20 Agricultural), giving
$M_{\text{frame}} = M_{\text{spar}}(1 + \gamma)$ with
$M_{\text{spar}} = C_{\text{body}}M_{\text{arms,eff}}$.

\subsection{Per-Class Structural Calibration}
\label{ssec:structcal}
Because folding-arm hinges, ruggedization hardware, and integration
scale differently across vehicle categories, a single set of physics
coefficients leaves a residual, category-correlated bias in predicted
empty mass.  AeroEval therefore applies a per-class structural
calibration multiplier $\lambda_{\text{class}}$ to the combined frame
and complexity mass, fitted on the held-out DIY cohort
(Section~\ref{ssec:calibration}):
\begin{equation}
\lambda_{\text{class}} = \begin{cases}
1.50 & \text{MicroAIO},\\
1.08 & \text{ConsumerFolding},\\
0.95 & \text{EnterpriseRugged (quad)},\\
0.80 & \text{EnterpriseRugged (hex/octo)}.
\end{cases}
\label{eq:lambda}
\end{equation}
These values are non-edge, gently regularized fits: MicroAIO racing
frames are systematically under-built by the generic scaling law, while
high-rotor-count enterprise frames are over-predicted; the multiplier
corrects the class mean without over-fitting individual platforms.

\subsection{Cantilever Resonance Verification}
\label{ssec:resonance}
To prevent vibrational resonance~\cite{blevins1979}, the arm natural
frequency must satisfy
$f_{\text{nat}} \geq 1.5\,f_{\text{bpf}} = 1.5\,N_b\,\text{RPS}_{\text{peak}}$,
where $N_b$ is blades per propeller (typically 2).  The
Rayleigh--Ritz/Dunkerley approximation is:
\begin{equation}
f_{\text{nat}} = \frac{1}{2\pi}\sqrt{\frac{3E_{\text{CF}}I_x}
{L_{\text{arm}}^3(M_{\text{tip}} + 0.24\mu L_{\text{arm}})}},
\label{eq:fnat}
\end{equation}
with $E_{\text{CF}} = 70$\,GPa and $\mu = \rho_{\text{CF}}A_{\text{wall}}$.

\subsection{Continuous Motor Scaling Laws}
\label{ssec:continuous-motor}
Prior to the discrete database search, Module~3 uses continuous scaling
laws to approximate motor properties.  The hover rotational speed is
$n_{\text{hover}} = \sqrt{M_{\text{to}}g/(N_r C_t\rho D_p^4)}$ with
$C_t \approx 0.12$, and $n_{\text{peak}} = n_{\text{hover}}/t_{\text{hover}}$
where $t_{\text{hover}} = \max(0.20, 1 - \text{margin}_{\text{thrust}})$.
The continuous target is $KV_{\text{target}} = 60\,n_{\text{peak}}/V_{\text{bus}}$
with $V_{\text{bus}} = V_{\text{full\_cell}}S$.  Winding resistance is
temperature-corrected,
$R_m(T) = R_{m,\text{nom}}(1 + \alpha_{\text{Cu}}(T - 25))$,
$\alpha_{\text{Cu}} = 0.00393\,\text{K}^{-1}$, and continuous total
motor mass scales with peak motor thrust,
$M_{\text{motor,tot}} = \max(0.008, N_r T_{\text{motor}}\cdot 0.00035\,C_{\text{motor}})$,
with $C_{\text{motor}} = 0.25$ for racing and $1.0$ otherwise.

\section{Electrochemical and Thermodynamic Modeling}
\label{sec:thermo}

\subsection{Battery Chemistry Profiles}
\label{ssec:chem}
Twelve lithium-family chemistry profiles are supported through
\texttt{chemistry\_profiles.json}.  Table~\ref{tab:chem} lists
Tremblay--Dessaint parameters for the six primary chemistries; six
additional consumer/enterprise/agricultural LiPo variants extend
coverage to optimized real-world pack specifications.

\begin{table}[ht]
\caption{Primary Battery Chemistry Profile Parameters.}
\label{tab:chem}
\centering
\scriptsize
\setlength{\tabcolsep}{2pt}
\begin{tabular}{lcccccc}
\toprule
Chemistry & $e_s$\,(Wh/kg) & $V_{\text{nom}}$\,(V) & $V_{\text{full}}$\,(V)
          & $B$ & $K$ & $R_i$\,(m$\Omega$) \\
\midrule
LiPo              & 180 & 3.70 & 4.20 & 5.5  & 0.046 & 5.0  \\
LiHV              & 195 & 3.80 & 4.35 & 10.0 & 0.040 & 4.0  \\
Li-ion NCA        & 260 & 3.60 & 4.20 & 12.0 & 0.040 & 20.0 \\
Li-ion NMC        & 240 & 3.60 & 4.20 & 8.0  & 0.060 & 15.0 \\
Solid State       & 400 & 3.60 & 4.20 & 15.0 & 0.030 & 35.0 \\
Li-S              & 500 & 2.10 & 2.45 & 5.0  & 0.100 & 50.0 \\
\bottomrule
\end{tabular}
\end{table}

\subsection{Tremblay--Dessaint OCV Formulation}
\label{ssec:tremblay}
Battery terminal voltage~\cite{tremblay2007, tremblay2009} is
$V_{\text{term}} = V_{\text{oc}}(\text{SOC}) - I\,R_{\text{sys}}$, with
open-circuit voltage:
\begin{equation}
V_{\text{oc}} = V_{\text{exp}} + (V_{\text{full}}-V_{\text{exp}})
  e^{-B_{\text{eff}}(1-\text{SOC})}
  - K\!\left(\frac{1}{\text{SOC}}-1\right),
\label{eq:voc}
\end{equation}
SOC clamped to $[0.02, 1.0]$ to prevent the polarization singularity at
depletion.  C-rate-dependent exponential-zone scaling is
$B_{\text{eff}} = B(1 + 0.18\ln C_{\text{rate}})$ for
$C_{\text{rate}} > 1$.  This logarithmic scaling extends the baseline
Tremblay--Dessaint model to high-discharge regimes typical of UAV
operation, where standard parameters extracted at low
C-rates cause voltage errors above 1C~\cite{cabello2015}.  The expanded
battery model of Cabello et~al.~\cite{cabello2015} splits the
exponential zone into two segments (NRMSE improvement 10--28\% at low
SOC); our single-parameter scaling achieves comparable accuracy without
increasing parameter count, keeping the per-invocation cost below
10\,ms.  P2D electrochemical models offer superior accuracy above
2C~\cite{kim2026} but require seconds per evaluation---prohibitive for
iterative sizing.

\subsection{Dynamic System Resistance}
\label{ssec:resistance}
\begin{equation}
R_{\text{sys}} = 10^{-3}R_{i,\text{base}}\cdot S
\cdot\frac{C_{\text{ref}}}{C_{\text{pack}}}
\cdot f_{\text{temp}}\cdot f_{\text{aging}}
+ \frac{R_{\text{ds,on}}}{N_r},
\label{eq:rsys}
\end{equation}
with $f_{\text{temp}} = e^{-0.013(T-25)}$ and
$f_{\text{aging}} = 1 + k_{\text{ag}}N_{\text{cyc}}$,
$k_{\text{ag}} \approx 0.001$--$0.0015$/cycle.

\subsection{Battery Packaging Factor}
\label{ssec:pkg}
The packaging multiplier $\eta_{\text{pkg}}$ in \eqref{eq:batmass}
models the dead-weight overhead of converting raw cells into a flight-ready
pack (bus bars, balance leads, housing, connectors).  Rather than a
single constant, it is computed per configuration.  For LiPo it starts
at a base of 1.04 and accrues additive terms: $+0.012$ per series cell
above 6S; $+0.06$ per additional parallel pack; a small-cell penalty up
to $+0.28\,(1 - C_{\text{pack}}/4000)$ for packs below 4000\,mAh
(reduced for consumer/micro classes); $+0.12$ for sub-250\,mm frames;
and $+0.10$ for enterprise packs below 14S.  Highly integrated
high-capacity enterprise packs ($\geq$14S) use 1.03, and cylindrical
Li-ion (NCA/NMC) and exotic chemistries use 1.15 to reflect cell
holders and fire-safety spacing.  Locked-capacity declarations are
interpreted as the \emph{total} system capacity at the pack terminals;
series packs do not multiply capacity, correcting a prior double-count
that produced physically impossible pack specific energies.

\subsection{ESC Mass and Wiring Harness}
\label{ssec:esc-mass-losses}
The ESC housing/heatsink mass is
$M_{\text{ESC}} = 0.00035\,I_{\text{ESC,rotor}}N_r C_{\text{esc}}S_{\text{esc}}$,
with $I_{\text{ESC,rotor}} = \max(15, I_{\text{ESC,rated}}/N_r)$ and
$C_{\text{esc}} = 0.10$ for racing/small frames, $1.0$ otherwise.  The
continuous down-scaling factor for light configurations is
$S_{\text{scale}} = 0.25$ for $M_{\text{ref}} < 0.3$\,kg, linearly
ramping to $1.0$ at $M_{\text{ref}} = 1.8$\,kg, with
$M_{\text{ref}} = M_{\text{frame}} + M_{\text{payload}}$.

The power-wiring harness mass is sized by the current each conductor
carries, not the pack total.  A multirotor's $N_r$ arm runs each carry
$I_{\text{max}}/N_r$ over a round-trip length $2L_{\text{arm}} = W$, so
the total copper mass is:
\begin{equation}
M_{\text{wiring}} = C_{\text{wire}}\cdot\frac{I_{\text{max}}}{N_r}
\cdot (N_r\cdot 2L_{\text{arm}})\cdot S_{\text{scale}}
= C_{\text{wire}}\,I_{\text{max}}\,W\,S_{\text{scale}},
\label{eq:mwiring}
\end{equation}
in which the rotor count cancels, with
$C_{\text{wire}} = 0.0003\,\text{kg/(A\,m)}$ for racing/small platforms
and $0.0015$ otherwise.  Billing the full pack current over all $N_r$
arm runs (rather than the per-arm branch current) over-counts copper by
a factor of $N_r$ and inflates hex/octo frames super-linearly; the
per-arm formulation in \eqref{eq:mwiring} removes this dimensional
inconsistency.  ESC parasitic switching loss is
$P_{\text{sw}} = \tfrac12 f_{\text{pwm}}C_{\text{oss}}V_{\text{oc}}^2$
with $f_{\text{pwm}} = 24{,}000$\,Hz and $C_{\text{oss}} = 2.5$--$5.0$\,nF.

\subsection{Voltage Collapse and Power Overflow Guard}
\label{ssec:vcollapse}
Current is solved from the quadratic power balance
$I = [V_{\text{oc}} - \sqrt{V_{\text{oc}}^2 - 4R_{\text{sys}}P_{\text{req}}}]/(2R_{\text{sys}})$.
If the discriminant is negative, \textbf{Voltage Collapse} is declared.
A numerical pathology arises when the cell count is auto-selected from a
preliminary low-mass estimate but the Banach loop converges to a
substantially heavier drone requiring more current on an
under-dimensioned bus; $P_{\text{req}}$ may overflow IEEE-754 double
precision before the discriminant check.  AeroEval guards against this
with an explicit check: if
$P_{e,\text{motor}} > 10^7$\,W or $P_{e,\text{motor}} \notin \mathbb{R}$,
Voltage Collapse is returned cleanly rather than propagating NaN
downstream.

\subsection{Transient Thermal Model}
\label{ssec:thermal}
Battery pack thermal dynamics are
$C_{\text{th,b}}\,dT_b/dt = I^2R_{\text{sys}} - hA_c(T_b - T_a)$, with
$C_{\text{th,b}} = M_{\text{bat}}C_p$, $C_p = 795$\,J/(kg\,K), and
scale-invariant convective cooling:
\begin{equation}
hA_c = h\cdot 6\!\left(\frac{M_{\text{bat}}}{2400}\right)^{2/3},
\quad h = 12\,\text{W/(m}^2\text{K)},
\label{eq:hacool}
\end{equation}
with $h = 120\,\text{W/(m}^2\text{K)}$ for racing (exposed high-speed
airflow).  If $T_b > 60^\circ$C, \textbf{Thermal Runaway} termination
is triggered.

\subsection{Path-Dependent Mass-Shedding}
\label{ssec:shedding}
For agricultural spraying, payload decreases continuously,
$M_{\text{pay}}(t) = M_{\text{pay},0} - \dot m\,t$ with
$\dot m = M_{\text{fluid}}/t_{\text{mission}}$.  For delivery, payload
drops discretely at $t_{\text{drop}} = r_{\text{drop}}\,t_{\text{total}}$.

\section{Discrete Hardware Optimization via Guided Grid Search}
\label{sec:opt}
The continuous optimum ($D_{\text{ideal}}$, $P_{\text{ideal}}$,
$KV_{\text{ideal}}$) serves as coordinates in an exhaustive search over
\texttt{cots\_database.json}.  The multi-objective $J$-score is
$J = w_1\Delta D + w_2\Delta KV + w_3\Delta p + \text{Pen}_a + \text{Pen}_\theta$
with normalized deviations $\Delta D$, $\Delta KV$, $\Delta p$ and
weights $w_1 = 100$, $w_2 = 10$, $w_3 = 50$.  The penalty terms are
$\text{Pen}_a = 1000\max(0, \theta_{\text{hvr}} - \theta_{\text{role}})$
and $\text{Pen}_\theta = 10\max(0, \theta_{\text{cr}} - 20^\circ)$.  If
the tested propeller violates \eqref{eq:overlap} or exceeds the
tip-Mach limit $V_{\text{tip}}/a > 0.75$, the branch is pruned
($J = 10^9$).

\section{Experimental Validation and Sensitivity Analysis}
\label{sec:valid}

\subsection{Non-Circular Calibrate/Test Protocol}
\label{ssec:protocol}
\label{ssec:calibration}
To eliminate the circularity of fitting sizing parameters on the same
data used for validation, three structural/packaging coefficients---the
per-class structural multiplier $\lambda_{\text{class}}$
\eqref{eq:lambda}, the battery packaging base, and the wiring
coefficient---were fitted exclusively on a held-out calibration cohort
of 20 DIY and legacy platforms
(Appendix~\ref{sec:appendix_calibration}).  The parameters were then
\emph{frozen} and evaluated blind on a disjoint cohort of 19 modern
commercial platforms (Table~\ref{tab:val20}).  This preserves a strict
calibrate/test separation: the reported commercial accuracy is a
held-out generalization result, not an in-sample fit.

\subsection{Commercial Blind-Test Cohort}
\label{ssec:20plat}
The complexity factor $k = \text{MTOW}_{\text{actual}}/\text{MTOW}_{\text{predicted}}$
quantifies unmodeled mechanical overhead.  Table~\ref{tab:val20} lists
the 19 commercial platforms.

\begin{table}[ht]
\caption{Commercial Blind-Test Validation Results (frozen parameters).}
\label{tab:val20}
\centering
\scriptsize
\setlength{\tabcolsep}{2.5pt}
\begin{tabular}{lcccc}
\toprule
Platform & Actual (kg) & Predicted (kg) & Error & $k$ \\
\midrule
  DJI Avata 2                      &  0.377 &  0.354 &   -6.1\% & 1.07 \\
  Parrot Anafi USA                 &  0.485 &  0.526 &   +8.5\% & 0.92 \\
  Custom 5-inch FPV                &  0.650 &  0.609 &   -6.3\% & 1.07 \\
  DJI Air 3                        &  0.720 &  0.785 &   +9.1\% & 0.92 \\
  DJI FPV Combo                    &  0.795 &  0.811 &   +2.0\% & 0.98 \\
  DJI Mavic 3 Classic              &  0.895 &  0.954 &   +6.6\% & 0.94 \\
  DJI Mavic 3 Enterprise           &  1.050 &  0.951 &   -9.4\% & 1.10 \\
  Custom 7-inch FPV                &  1.200 &  1.192 &   -0.7\% & 1.01 \\
  DJI Phantom 4 Pro V2.0           &  1.375 &  1.192 &  -13.3\% & 1.15 \\
  Teal Golden Eagle                &  1.400 &  1.319 &   -5.8\% & 1.06 \\
  Skydio X10                       &  2.490 &  2.444 &   -1.8\% & 1.02 \\
  DJI Inspire 3                    &  3.995 &  3.571 &  -10.6\% & 1.12 \\
  DJI Inspire 2                    &  4.250 &  4.139 &   -2.6\% & 1.03 \\
  Sony Airpeak S1                  &  4.400 &  4.517 &   +2.7\% & 0.97 \\
  DJI Matrice 300 RTK              &  9.000 &  8.373 &   -7.0\% & 1.07 \\
  DJI Matrice 350 RTK              &  9.200 &  8.349 &   -9.3\% & 1.10 \\
  Watts Innovations Prism          & 20.360 & 16.505 &  -18.9\% & 1.23 \\
  DJI FlyCart 30                   & 40.000 & 35.015 &  -12.5\% & 1.14 \\
  DJI Agras T30                    & 76.000 & 73.328 &   -3.5\% & 1.04 \\
\bottomrule
\end{tabular}
\end{table}

Across the 19 commercial platforms (377\,g to 76\,kg),
the engine achieves a MTOW MAPE of \textbf{7.2\%} (95\% CI:
5.1\%--9.3\%), median absolute error 6.6\%,
$\text{RMSE}_{\text{MTOW}} = 1.59$\,kg, mean complexity factor
$k \approx 1.05 \pm 0.08$, and $k$-range $0.92$--$1.23$.  Battery mass
is predicted within 7.9\% MAPE on the 16 single-pack platforms.  The
four redundant multi-battery enterprise platforms (Inspire~2/3,
Matrice~300/350) exhibit a systematic battery-mass underprediction of
    $\approx$50\%: the engine sizes energy capacity, whereas these
platforms carry dual hot-swappable packs sized for redundancy rather
than minimum mission energy.  This is a disclosed model boundary rather
than a fitting error, and does not materially affect MTOW because the
redundant pack mass is partly offset by the vehicles' higher structural
integration.

\subsection{Case Studies}
\label{ssec:cases}
Table~\ref{tab:cases} summarizes six role-specific case study outputs.

\begin{table}[ht]
\caption{Representative Case Study Results (Six Role Profiles).}
\label{tab:cases}
\centering
\scriptsize
\setlength{\tabcolsep}{2pt}
\begin{tabular}{lccccc}
\toprule
Role / Platform & MTOW (kg) & Prop & KV & Battery & Conv.$^{\text{a}}$ \\
\midrule
Search \& Rescue Quad (Imaging) & 5.57  & 26$\times$9     & 200  & 4S 12650\,mAh  & --- \\
Delivery Quad (Delivery)        & 12.41 & 39.5$\times$14  & 80   & 6S 36861\,mAh  & --- \\
Agricultural Spray Octo (Smart) & 50.33 & 38$\times$12    & 50   & 12S 14568\,mAh & --- \\
FPV Racing Quad                 & 0.31  & 5$\times$2      & 2300 & 4S 1264\,mAh   & --- \\
LiDAR Mapping Quad              & 2.81  & 21$\times$7     & 200  & 3S 7071\,mAh   & --- \\
Inspection Quad (Coaxial)       & 7.77  & 15$\times$4     & 450  & 6S 26881\,mAh  & --- \\
\bottomrule
\end{tabular}
\vskip 2pt
{\scriptsize $^{\text{a}}$ Convergence counts are not reported for
role runs using component locks that bypass the active mass-coupling
loop; convergence is satisfied by construction.}
\end{table}

\subsection{Agricultural and Delivery Path-Dependent Sizing}
\label{ssec:agri}
An agricultural sizing case study inspired by the DJI Agras T30 mission
profile (20\,kg liquid payload) was optimized under both static and
dynamic mass-shedding models.  The dynamic model achieves a
\textbf{33.7\% energy capacity saving} and \textbf{40.6\% structural
frame mass saving} relative to static 20\,kg sizing, with MTOW
$= 35.27$\,kg (static baseline 48.48\,kg).  A delivery role (3.5\,kg
parcel, $r_{\text{drop}} = 0.50$) achieves a 27.1\% reduction in
average takeoff mass.

\subsection{14-Parameter Sensitivity Analysis}
\label{ssec:sensitivity}
Table~\ref{tab:sens} reports the sensitivity of predicted MTOW and
hover flight time to 14 independently swept parameters across $>$300
simulations.  No NaN or non-convergent runs were observed.

\begin{table}[ht]
\caption{14-Parameter Sensitivity Analysis Summary.}
\label{tab:sens}
\centering
\scriptsize
\setlength{\tabcolsep}{2pt}
\begin{tabular}{lcc}
\toprule
Parameter & MTOW Sensitivity & Flight Time Sensitivity \\
\midrule
Payload mass               & $+1.692$\,kg/kg       & $-1.703$\,min/kg \\
Propeller diameter limit   & $+0.080$\,kg/in       & $+0.011$\,min/in \\
Forward speed (to 25.5\,m/s) & $+0.062$\,kg/(m/s)    & $-0.691$\,min/(m/s) \\
Forward speed ($>$25.5\,m/s) & \textit{diverges}    & \textit{diverges} \\
Climb speed                & $+0.128$\,kg/(m/s)    & $-0.461$\,min/(m/s) \\
Altitude (0--5000\,m)      & $+0.030$\,kg/1000\,m  & $-0.200$\,min/1000\,m \\
Ambient temp ($+10^\circ$C) & $+0.004$\,kg/$10^\circ$C & $+0.037$\,min/$10^\circ$C \\
Aux power (+10\,W)         & $+0.200$\,kg/10\,W    & $-0.385$\,min/10\,W \\
Added drag area            & $+0.062$\,kg/0.1\,m$^2$ & $-2.531$\,min/0.1\,m$^2$ \\
Center of gravity shift    & $0.000$\,kg/5\,cm     & $-0.169$\,min/5\,cm \\
Battery capacity lock      & $+0.100$\,kg/1000\,mAh & $+0.100$\,min/1000\,mAh \\
Propeller diameter lock    & $-0.150$\,kg/in       & $0.000$\,min/in \\
Delivery drop mass         & $-0.022$\,kg/kg       & $0.000$\,min/kg \\
Delivery drop time ratio   & $+0.033$\,kg/0.1\,ratio & $0.000$\,min/0.1\,ratio \\
Spray rate                 & $-0.490$\,kg/(0.005\,kg/s) & $0.000$\,min/(0.005\,kg/s) \\
\bottomrule
\end{tabular}
\end{table}

\section{Numerical Robustness and Test Verification}
\label{sec:testing}
A 28-case physics validation suite exercises monotonicity, boundary
conditions, and structural sanity: output-schema completeness across
all JSON fields; physics bounds (T/W in $[1.5, 8.0]$, peak battery
temperature in $[20, 60]^\circ$C, cell count in $[2, 12]$, battery mass
fraction in $[10\%, 75\%]$); monotonicity (payload$\uparrow$
$\Rightarrow$ MTOW$\uparrow$; 5000\,m MTOW $>$ sea-level; ambient
temperature$\uparrow$ $\Rightarrow$ peak battery temperature$\uparrow$;
aux power$\uparrow$ $\Rightarrow$ shorter approved flight time); mission
coverage (hover-only, forward-only, coaxial X8, 5000\,m, NMC vs.\ LiPo);
and thermal consistency.  A 74-case full-schema coverage suite
systematically exercises every field in the 57-parameter input contract
across nine sections (payload roles, airframe classes, all twelve
chemistries, lock fields, aerodynamic overrides, structural overrides,
battery options, flight toggles, and output completeness).  Under the
calibrated model all 102 tests pass with zero unexpected failures across
$>$500 engine invocations.  Two engine defects were discovered and fixed
during this campaign: (1)~a power-overflow in \texttt{Thermodynamics.cpp}
when a locked propeller is undersized relative to the payload
(Section~\ref{ssec:vcollapse}); and (2)~an incorrect racing-role thrust
margin, corrected to 80\% in Table~\ref{tab:roles}.

\section{Limitations and Future Work}
\label{sec:limits}
The current engine boundaries are: (i)~\textbf{rigid-disk actuator
aerodynamics}, omitting aeroelastic blade flapping, flutter, and
tip-deflection losses (4--7\% for propellers $>24$\,in above 3000\,RPM);
(ii)~\textbf{lumped unmodeled mass}---folding-arm hinges, IP-rated
seals, and connector housings (200--600\,g on enterprise rigs) are
absorbed by the class multiplier $\lambda_{\text{class}}$ rather than
modeled from first principles; (iii)~\textbf{redundant multi-battery
platforms} are sized for mission energy, not pack redundancy, causing
the disclosed $\approx$50\% battery-mass underprediction on such
vehicles; (iv)~\textbf{quasi-static, deterministic dynamics} with no
stochastic wind fields or transient gust response; and
(v)~\textbf{intermediate-altitude Banach attractors} (1000--4000\,m)
where the corrected seed may not uniquely determine the global minimum,
occasionally producing $\pm 6\%$ local non-monotonicity.  Planned
extensions include a blade-element momentum rotor model, multi-attractor
detection via Jacobian eigenvalue analysis, an explicit redundant-pack
sizing mode, motor-thermal feedback into the convergence loop,
Monte-Carlo gust-margin sizing, and fixed-wing/tilt-rotor support.

\section{Conclusion}
\label{sec:conclusion}
AeroEval presents a formally convergent, multidisciplinary design
optimization engine for COTS multirotor UAVs, implemented as a
cross-platform C++17 executable and exposed through a dual-mode web
application.  The solver resolves the mutual coupling between structural
mass, battery electrochemistry, aerodynamic performance, and kinematics
in a single unified Banach fixed-point loop, avoiding the Mass Snowball
divergence of sequential single-domain solvers.  Under a strict,
non-circular calibrate/test protocol, the engine achieves \textbf{7.2\%
MTOW MAPE} on a 19-platform commercial blind-test cohort spanning 377\,g
to 76\,kg ($k \approx 1.05 \pm 0.08$;
$\text{RMSE}_{\text{MTOW}} = 1.59$\,kg), with battery mass within 7.9\%
on single-pack platforms.  The path-dependent mass-shedding model
reduces sizing conservatism by up to 33.7\% for agricultural sprayers.
Two qualitatively distinct interface modes---the Mission Workspace and
the Physics Workspace---serve non-expert and expert practitioners from
the same underlying physics, and a 102-case verification suite confirms
numerical robustness across all 12 chemistry profiles, 6 role types, 4
airframe classes, and all 57 input fields.  The primary remaining
accuracy gaps---unmodeled enterprise mechanical overhead and redundant
multi-battery sizing---are well-characterized by the $k$-factor taxonomy
and amenable to future component-level accounting.

\appendices
\section{Calibration Set Characterization}
\label{sec:appendix_calibration}
The held-out calibration cohort comprises 20 DIY and legacy platforms
spanning 580\,g racing quads to 18\,kg octocopters
(Table~\ref{tab:calibration_set}).  With the frozen coefficients this
cohort yields a MTOW MAPE of 26.1\% (95\% CI: 18.5\%--33.7\%), median
25.3\%, and a mean complexity factor $k \approx 1.03 \pm 0.35$.  The
markedly higher variability---relative to the commercial test cohort's
7.2\%---reflects the non-integrated carbon-fiber builds, heterogeneous
wiring, and hand-assembled component choices typical of experimental
platforms, and is exactly why this cohort is used for calibration
rather than headline reporting: fitting to its scatter would overstate
achievable accuracy on production hardware.

\begin{table}[ht]
\caption{Calibration Dataset and Held-out Sizing Results.}
\label{tab:calibration_set}
\centering
\scriptsize
\setlength{\tabcolsep}{3pt}
\begin{tabular}{lccc}
\toprule
Platform & Actual (kg) & Predicted (kg) & Error \\
\midrule
  Holybro QAV250            &  0.580 &  0.333 & -42.5\% \\
  Holybro QAV-R 5"          &  0.650 &  0.335 & -48.5\% \\
  DJI FlameWheel F450       &  1.200 &  0.973 & -19.0\% \\
  3DR Iris+                 &  1.282 &  1.186 &  -7.5\% \\
  Holybro S500 V2           &  1.450 &  1.255 & -13.5\% \\
  3DR Solo                  &  1.800 &  1.238 & -31.2\% \\
  DJI FlameWheel F550       &  2.100 &  1.907 &  -9.2\% \\
  Custom 9-inch Long Range  &  2.200 &  1.652 & -24.9\% \\
  Tarot 650 Sport           &  2.800 &  3.520 & +25.7\% \\
  Custom 10-inch Cinelifter &  3.800 &  3.971 &  +4.5\% \\
  Tarot FY680               &  4.500 &  5.449 & +21.1\% \\
  Custom 13-inch Heavy Quad &  5.500 &  5.919 &  +7.6\% \\
  Freefly Astro Max         &  7.600 &  8.462 & +11.3\% \\
  DJI S900                  &  8.200 & 10.438 & +27.3\% \\
  Tarot T960                &  8.500 & 11.586 & +36.3\% \\
  Hexadrone Tundra          &  8.500 &  9.082 &  +6.9\% \\
  DJI S1000                 & 10.000 & 14.724 & +47.2\% \\
  Aurelia X6 Pro V2         & 10.316 & 17.614 & +70.7\% \\
  Tarot Iron Man 1000       & 12.500 & 16.942 & +35.5\% \\
  Tarot T18                 & 18.000 & 23.595 & +31.1\% \\
\bottomrule
\end{tabular}
\end{table}

\end{document}